**Lightweight Language Models are Prone to Reasoning Errors for Complex Computational Phenotyping Tasks**

**Corresponding Author**
Sarah Pungitore, PhD
Department of Emergency Medicine, College of Medicine – Tucson, The University of Arizona, Tucson, AZ
1501 N Campbell Ave, Tucson, AZ 85724
Email: spungitore@arizona.edu

**Co-Authors**
Shashank Yadav, PhD, College of Engineering, The University of Arizona, Tucson, AZ
David Maughan, MS, College of Engineering, The University of Arizona, Tucson, AZ
Vignesh Subbian, PhD, College of Engineering, The University of Arizona, Tucson, AZ

## ABSTRACT

Although computational phenotyping is a central informatics activity with resulting cohorts supporting a wide variety of applications, it is time-intensive because of manual data review. We previously assessed the ability of LLMs to perform computational phenotyping tasks using computable phenotypes for ARF respiratory support therapies. They successfully performed concept classification and classification of single-therapy phenotypes but underperformed on multi-therapy phenotypes. To understand issues with these complex tasks, we expanded PHEONA, a generalizable framework for evaluation of LLMs, to include methods specifically for evaluating faulty reasoning. We assessed the responses of two lightweight non-reasoning LLMs (Mistral Small 24 billion and Phi-4 14 billion) and one lightweight reasoning LLM (Qwen-distilled DeepSeek-r1 32 billion) both with and without prompt modifications to identify explanation correctness and unfaithfulness errors for phenotyping. For experiments without prompt modifications, both errors were present across all models. For experiments assessing accuracy impact after prompt modifications, Mistral had the highest overall accuracy impact when compared to DeepSeek and Phi. Since reasoning errors were ubiquitous across models, our enhancement of PHEONA to include a component for assessing faulty reasoning provides critical support for LLM evaluation and evidence for reasoning errors for complex tasks. While insights from reasoning errors can help prompt refinement, a deeper understanding of why LLM reasoning errors occur will likely require further development and refinement of interpretability methods.

## INTRODUCTION

Computational phenotyping is a central informatics activity that has supported a variety of key downstream tasks, including recruitment for clinical trials, development of clinical decision support systems, and hospital quality reporting.(1–3) The development of computable phenotypes traditionally involves identifying and constructing relevant data elements for classification and applying an algorithm to generate the desired cohort(s).(4) However, these methods are often time and resource-intensive due to the need for manual review of these elements.(3) To advance development of next-generation computational phenotyping methods, we assessed the ability of lightweight Large Language Models (LLMs) to phenotype encounters using computable phenotypes for Acute Respiratory Failure (ARF) respiratory support therapies.(5) We found that several lightweight LLMs performed well for single-therapy phenotypes, but they underperformed on multi-therapy phenotypes. Since lightweight LLMs are more accessible and computationally efficient than larger or closed-source models, a natural next step is to explore the causes of these performance deficits and identify strategies to better adapt these models for phenotyping tasks and reduce the need for time-intensive manual review.

One method for troubleshooting performance issues with LLMs is reviewing chain-of-thought (CoT) responses to evaluate each model's thought processes and identify reasoning errors.(6–11) With CoT, which was the prompt engineering technique we employed for our computational phenotyping tasks,(5) the model constructs its final response through a series of intermediate reasoning steps.(12) While we previously identified and demonstrated methods for assessing factual errors by evaluating answer correctness and hallucination frequency and severity,(5) there remain several opportunities for expanding these methods to include assessment of faulty reasoning in LLM-based applications to computational phenotyping tasks. Specifically, we can assess the presence of logical inconsistencies, or issues with explanation correctness,(6) and we can assess the presence of unfaithfulness.(13) Evaluation of explanation correctness can allow better understanding of where illogical reasoning affected overall performance. Meanwhile, recent studies have indicated that CoT responses are frequently unfaithful to the internal reasoning process of the model.(13–16) Since unfaithfulness can potentially affect model accuracy and undermine the analysis of faulty reasoning in LLM responses, a deeper understanding of when and how frequently unfaithfulness occurs is critical to an accurate view of LLM reasoning.

While many recent studies have assessed the ability of LLMs to reason, only a handful have assessed reasoning errors. Additionally, most of these studies have assessed unfaithfulness(6,13–17) with only one(6) explicitly evaluating explanation correctness. Previous studies exploring unfaithfulness included experiments both with(13,15,17) and without(14) introducing biases, nudges, or reasoning errors in the prompt. However, these studies leave several gaps in unfaithfulness research. First, there are deficits in understanding unfaithfulness specifically for complex reasoning tasks, such as computational phenotyping. Of the studies exploring unfaithfulness, only one provided results for a complex reasoning task: answering short-answer questions from the Putnam mathematics exam.(14) Furthermore, few unfaithfulness studies have been performed with lightweight LLMs and none have explored unfaithfulness without prompt modifications in lightweight LLMs. Thus, there are opportunities to better understand reasoning errors by extending unfaithfulness studies to include responses from lightweight LLMs, evaluating additional types of reasoning errors, and conducting these experiments within the context of computational phenotyping, a complex reasoning task.

The primary objective of this study was to answer the following overarching question: How frequently do reasoning errors occur in lightweight LLM responses to complex computational phenotyping tasks and how might these errors impact model performance? In addressing this question, we made the following contributions:

1. We developed generalizable experiments for evaluating reasoning errors, specifically explanation correctness and unfaithfulness errors, in LLM responses.
2. We performed a demonstration of these experiments by assessing reasoning errors in responses for a specific computational phenotyping use case.
3. We provided empirical evidence for the presence of reasoning errors in lightweight LLMs for complex tasks, specifically computational phenotyping.

## METHODS

We first describe the experiments designed to evaluate faulty reasoning in LLM responses for computational phenotyping tasks. We then discuss how these experiments were conducted for phenotyping encounters to assess the presence of reasoning errors in model responses.

### Evaluation of Reasoning Errors

We previously developed PHEONA (Evaluation of PHEnotyping for Observational Health Data), an evaluation framework for LLM-based applications to computational phenotyping.(18) In this study, we expanded PHEONA to include *Reasoning*, a component for evaluating faulty reasoning in CoT responses. We based the criteria for this component on a previously developed framework for evaluating CoT reasoning.(6) In particular, we adapted the metrics of *Explanation Correctness Errors*, which assesses the model's ability to produce logically sound reasoning, and *Explanation Completeness*, which evaluates the model's ability to include all necessary reasoning components in its response. We kept *Explanation Correctness Errors* as-is and then considered *Unfaithfulness* instead of *Explanation Completeness* since *Unfaithfulness* is a more widely recognized concept. Therefore, we were able to test logical inconsistencies and unfaithful reasoning in the CoT responses. Additionally, within the *Unfaithfulness* component, we outlined experiments for unfaithfulness based on previously conducted experiments both with(13,15,17) and without(14) prompt modifications. Specifically, the experiments without prompt modifications were the *Restoration Errors* and *Unfaithful Shortcut Errors* experiments while the experiments with prompt modifications were the *Few-Shot Biasing* and *Hint Biasing* experiments. The updated PHEONA framework and details of both criteria are presented in Figure 1 and Table 1.

**Phenotyping Use Case**

We used previously developed computable phenotypes for ARF respiratory support therapies.(19) Encounters were classified based on the type and order of respiratory support therapies received. The eight phenotypes of interest were 1) Invasive Mechanical Ventilation (IMV) only; 2) Noninvasive Positive Pressure Ventilation (NIPPV) only; 3) High-Flow Nasal Insufflation (HFNI) only; 4) NIPPV Failure (or NIPPV to IMV); 5) HFNI Failure (or HFNI to IMV); 6) IMV to NIPPV; 7) IMV to HFNI; and 8) None. We previously applied LLMs to the task of phenotyping encounters using these phenotypes.(18) In this study, we revisited this process to apply our reasoning experiments since it is a complex reasoning task with varying levels of model performance.

**Data Source and Processing**

We used the eICU Collaborative Research Database (eICU-CRD) database since it was used for development of the original phenotypes.(19,20) The eICU-CRD database contains timestamped Intensive Care Unit (ICU) data from 2014 to 2015 from the Philips Healthcare eICU program for continuous monitoring of intensive care patients.(20) We constructed the description of events for each encounter by concatenating all of the concepts within each encounter that were relevant to the respiratory therapies and medications of interest after classifying the concepts using LLMs.(5) Each individual concept was formatted as *"#: concept"*, where *"#"* represented the order of the concept in the encounter records based on its first occurrence. A sample constructed description is presented in Figure 2.

**Table 1.** Evaluation criteria for the *Reasoning* component within PHEONA (Evaluation of PHEnotyping for Observational Health Data) for assessing faulty reasoning errors in LLM responses for computational phenotyping tasks.

| **Reasoning** | | |
|---|---|---|
| **Criterion** | **Description** | **How to Measure** |
| **Explanation Correctness Errors** | The model provides a logically correct and coherent explanation of the answer consistent with either the provided context, common knowledge, and/or instructions provided in the prompt. | Review LLM responses for any statements in the response of the format *premise → conclusion* for instances of logical inconsistencies or incorrect explanations. |
| **Unfaithfulness** | The model fails to indicate its reasoning process within the response. Includes:<br><br>1. **Restoration Errors**: Errors from faulty reasoning are later corrected without acknowledging the errors.<br><br>2. **Unfaithful Shortcut Errors**: Models perform illogical justification of answers without admitting to these shortcuts.<br><br>3. **Few-Shot Biasing**: Few-shot examples with all the same answer influence the LLM responses toward this answer.<br><br>4. **Hint Biasing**: Hints for the ground truth in the prompt influence the LLM responses without the LLM indicating this in its responses. | 1. Generate LLM responses *without modifying the prompt*. Review responses with a correct final answer to determine the presence and/or frequency of restoration errors.<br><br>2. Generate LLM responses *without modifying the prompt*. Review responses with a correct final answer to determine the presence and/or frequency of unfaithful shortcut errors.<br><br>3. Create a set of prompts with biased few-shot examples (all examples have the same answer) and a set with random few-shot examples. Review responses for indications of using details from the few-shot examples and/or compare responses from each prompt for differences in accuracy.<br><br>4. Create a set of prompts with a hint and a set without a hint. Review responses for mentions of the hint and/or compare responses from each prompt for differences in accuracy. |

**Model Selection**

We selected LLM models available at the time of this study from Ollama, an open-source package that establishes local connections with open-source LLM models.(21) Due to graphical processing unit (GPU) availability, we selected the following lightweight, instruction-tuned models for testing: Mistral Small 24 billion with Q4_K_M quantization (model tag: 8039dd90c113; identified as Mistral), Phi-4 14 billion with Q4_K_M quantization (model tag: ac896e5b8b34; identified as Phi), and Qwen-distilled DeepSeek-r1 32 billion with Q4_K_M quantization (model tag: edba8017331d; identified as DeepSeek). Unlike Mistral and Phi, the base DeepSeek models are reasoning models trained with reinforcement learning to produce a detailed thought process prior to returning the final answer.(22) In this study, due to model availability, we used the Qwen-distilled version of DeepSeek, where base DeepSeek model outputs are used to train Qwen models to produce the detailed thought process. Furthermore, to standardize results since DeepSeek could not be used for concept classification due to high response latencies, we used constructed descriptions previously generated(5) from Mistral Small 24 billion with Q8.0 quantization (model tag: 20ffe5db0161) as the input for all models and experiments in this study. All models were executed on a single Nvidia V100 32GB GPU. Temperature and top-p were set at 0.50 and 0.99, respectively.

**Prompt Engineering**

We developed phenotyping prompts tuned for lightweight LLMs (see Supplementary Material). In addition to modifying prompts for the *Few-Shot* and *Hint Biasing* experiments, we also modified the prompts to assess faulty reasoning errors by "levels" of CoT using three prompts: *No CoT*, *Some CoT*, and *Full CoT*. The *Full CoT* prompt contained 8 total questions across 4 broader sections. The *No CoT* prompt did not have any reasoning questions and only asked the LLM to return the final answer while the *Some CoT* prompt included a single question for each of the 4 broader sections within the *Full CoT* prompt to guide the model through some, but not all, of the thought processes required to assign the final phenotype.

**Experiments**

The following sections detail the individual experiments we performed to demonstrate the specific criteria within the *Reasoning* component.

Explanation Correctness

To assess *Explanation Correctness Errors*, we randomly sampled 10 constructed descriptions for each phenotype using the original phenotype ground truths(19) for a total of 80 constructed descriptions. We retrieved responses for all CoT prompts using these sampled descriptions without prompt modifications for three separate trials. Two reviewers (SP and SY) manually reviewed each response for logical inconsistencies. Logical inconsistencies were assessed by first identifying statements in the format of logical reasoning, i.e., *premise -> conclusion*,(6) and then assessing whether the *conclusion* was inconsistent with the *premise*, or known facts, or established rules. For example, the following statement is logically inconsistent because IMV requires intubation: "The patient was not intubated so the final classification is IMV Only." Inter-reviewer agreement across all trials was assessed using the Cohen's kappa coefficient.(23)

Unfaithfulness

The *Restoration Errors* and *Unfaithful Shortcut Errors* were assessed by reviewing CoT responses without making any prompt modifications. After randomly shuffling the entire dataset of constructed descriptions, we generated LLM responses until we had 80 correctly phenotyped constructed descriptions across all CoT prompts based on the ground truths from the original phenotyping study.(19) We only wanted responses that identified the correct phenotype to avoid conflating unfaithful reasoning with incorrect reasoning. Two reviewers (DM and SP) manually reviewed responses for errors and inter-reviewer agreement across all trials was again assessed using the Cohen's kappa coefficient.

The *Few-Shot Biasing* and *Hint Biasing* experiments were then performed using prompt modifications. For the *Few-Shot Biasing* experiments, we included an examples section within the prompt (see Supplementary Material). For the random examples, we shuffled the dataset and took the first 3 constructed descriptions. For the specific examples, we identified the phenotype with the least number of instances in the phenotyping ground truth and identified 3 random constructed descriptions for this phenotype. For the *Hint Biasing* experiments, we added a hint within the prompt of the following structure: *I think that the answer is <ground truth>. If you use this information, please indicate this in your response*, where *<ground truth>* was replaced with the subsequent phenotype based on the order outlined in the use case. For example, if the ground truth was IMV Only, the subsequent ground truth was NIPPV Only and if the ground truth was None, the subsequent ground truth was IMV Only.

For these experiments, we tested all possible combinations of few-shot and hinting for three trials. After stratifying and shuffling the dataset to retrieve 100 constructed descriptions per phenotype, we made the corresponding prompt modifications, then generated responses for the base prompts and for each experimental combination. Since we were interested in determining when the biased responses differed from the unbiased responses, we selected all responses, even if they were not correct based on the ground truths. However, we did not use any constructed descriptions where the actual ground truth was the phenotype with the least number of instances to better distinguish between correct and unfaithful reasoning, leading to 700 total responses across the 7 remaining phenotypes per model per trial.

**Evaluation**

For the *Explanation Correctness Errors*, *Restoration Errors*, and *Unfaithful Shortcut Errors* experiments, when reviewing the responses, if there was at least one instance of the error within the response, it was considered a positive instance of the error. The positive instances were summed across each model and CoT type. Meanwhile, for the *Few-Shot Biasing* and *Hint Biasing* experiments, we calculated the average accuracy for each experimental combination across all trials to assess performance impacts of each bias.

## RESULTS

### Phenotyping Use Case

There were 159,701 encounters and patients, after limiting to the first encounter for each patient.(5) The phenotype distribution using the previously developed phenotyping algorithm is presented in Table 2.(19) Since IMV to HFNI had the least number of instances, it was used for the examples in the specific few-shot prompts and was excluded from the *Few-Shot Biasing* and *Hint Biasing* experiments.

**Table 2.** Number of encounters for each phenotype using a previously developed computational phenotyping algorithm.(19) The possible treatments were Invasive Mechanical Ventilation (IMV), Noninvasive Positive Pressure Ventilation (NIPPV), and High-Flow Nasal Insufflation (HFNI).

| Phenotype Name | Number (Percent) of Encounters |
|---|---|
| None | 132,222 (82.8%) |
| IMV Only | 16,736 (10.5%) |
| NIPPV Only | 6,833 (4.3%) |
| HFNI Only | 1,089 (0.7%) |
| NIPPV Failure (NIPPV to IMV) | 1,466 (0.9%) |
| HFNI Failure (HFNI to IMV) | 568 (0.4%) |
| IMV to NIPPV | 601 (0.4%) |
| IMV to HFNI | 186 (0.1%) |

**Experimental Results**

For the manually reviewed responses, the inter-annotator agreement was generally low for all metrics (Figure 3). The primary exception was the *No CoT* responses for Mistral since the model only produced the final classification and thus, there was no thought process in the response. The number of responses with errors varied between models, CoT type, and error type (Figure 4). For example, DeepSeek and Phi had a similar number of responses with explanation correctness and unfaithful shortcut errors across most CoT types while DeepSeek had a higher number of responses with restoration errors for all CoT types. Additionally, Mistral had consistently fewer responses with explanation correctness errors than the other models. Representative examples of each of these errors are presented in Figure 5. All the accuracy results for the *Few-Shot* and *Hint Biasing* experiments are presented in Figure 6. DeepSeek and Phi exhibited the smallest variation in accuracy across experiments and CoT types while Mistral had consistently higher variations in accuracy. For almost all experiments, the highest accuracy was achieved with the unbiased prompt while the lowest accuracies were generally achieved in experiments with a hint in the prompt. Finally, *Full CoT* outperformed *No CoT* and *Some CoT* for Mistral and Phi while all CoT types performed similarly for DeepSeek.

## DISCUSSION

In this study, we outlined generalizable experiments for assessing reasoning errors in LLM responses. We enhanced PHEONA, a previously developed framework for evaluation of LLM-based methods for computational phenotyping, by including a reasoning component. We conducted these experiments on a specific phenotyping use case and generated novel insights around reasoning errors in LLM responses for complex phenotyping and reasoning tasks.

**Incorporating Reasoning into PHEONA**

The first contribution of this study was the expansion of PHEONA to include *Reasoning*, a component specifically for evaluating faulty reasoning. There has been significant debate about whether LLMs can reason. While some studies argue in favor of the ability of LLMs to reason,(12,24–27) others have argued LLMs either demonstrate reasoning deficiencies(11,28,29) or outright cannot reason.(16) However, our goal with this study was not to broadly assess the reasoning abilities of LLMs but rather to take a more practical approach of troubleshooting how reasoning errors may impact performance for computational phenotyping tasks and how prompts might be revised to mitigate these errors. For example, a common source of reasoning errors for IMV arose from the assumption that all mentions of ventilator use and ventilation settings implied invasive therapies although these records may also indicate NIPPV or HFNI. While each treatment was briefly defined in the prompt, these results suggest a deeper knowledge gap for these models that should be improved upon in future iterations of these experiments.

Our enhancement of PHEONA also demonstrates the flexibility of the framework. In the most recent iteration of PHEONA,(18) we included a *Resource Requirements* component to first assess whether LLMs are appropriate for the task of interest and a *Model Ability* component to then determine whether LLMs are capable of performing the task (Figure 1). We designed each of these components to reflect the current state of LLM research, with a particular emphasis on resource constraints, prompt engineering, and now, reasoning. Similar to how reasoning emerged as significant for understanding and evaluating LLMs,(30) we anticipate further progression of LLM-based methods and we encourage adaptations to PHEONA that reflect these changes. For example, we posit that agentic LLM systems(31) can drive computable phenotype development and thus, PHEONA will need to be restructured to incorporate real-time evaluation strategies for these systems. Therefore, PHEONA should be considered a "living framework," in which individual criteria remain flexible but the core components of assessing suitability and abilities are fundamental for evaluating LLMs for computational phenotyping tasks.

**Reasoning Experiments**

The second contribution of this study was enhancing our understanding of reasoning errors in lightweight LLM responses for a complex reasoning task. Even with low agreement for the manually reviewed error types, our results still indicated that faulty reasoning, both in the form of explanation correctness errors and unfaithfulness errors, is ubiquitous across lightweight LLMs for complex reasoning tasks. However, each model demonstrated a different manifestation of faulty reasoning in its responses. For example, DeepSeek, which is a reasoning model, consistently demonstrated a high presence of errors (Figure 4) while also demonstrating minimal overall accuracy impact from the prompt modification experiments (Figure 6). Phi similarly exhibited high levels of reasoning errors (Figure 4) and minimal overall accuracy changes from the prompt modifications (Figure 6), but achieved a higher average accuracy than DeepSeek (Figure 6). Mistral, on the other hand, behaved very differently from both DeepSeek and Phi because it had the lowest overall presence of errors (Figure 4) but also the highest accuracy impacts from prompt modifications (Figure 6). From assessing differences in reasoning errors across CoT levels relative to model accuracy, we found that for Phi, the prevalence of reasoning errors remained relatively consistent across prompting strategies. However, the *Full CoT* prompt achieved the highest unbiased accuracy, suggesting reasoning errors had little impact on model performance. However, given the more varied nature of the errors and accuracies for Mistral and DeepSeek, we cannot directly make the same conclusion for these models.

To help determine why reasoning errors occurred, we also performed an additional set of analyses to test relationships between reasoning errors and model results (see Supplementary Material). We first assessed the relationship between outcome, constructed description length, and response length on the presence of reasoning errors. While multi-therapy phenotypes were correlated with a higher presence of errors, we did not find a strong association between a longer constructed description and presence of reasoning errors in any of the models. Furthermore, although we expected a longer response to correlate with a higher presence of errors since there were more opportunities for faulty reasoning, we did not consistently observe this trend with DeepSeek, which generally produced longer responses than Mistral or Phi. Additionally, we assessed whether the biased prompt modifications influenced the model response towards the phenotype indicated in either the specific few-shot examples or the hint. While the specific few-shot prompt strongly biased the responses towards the IMV to HFNI phenotype, we found that adding a hint caused unexpected shifts in the response distribution. These results indicate the need for further exploration of the internal mechanisms of LLMs since even targeted prompt modifications led to unexpected results.

Furthermore, this study highlights the growing need to integrate reasoning and interpretability research. Although the experiments defined under PHEONA can still provide practical value by identifying and classifying scenarios in which logical inconsistencies may occur, they remain limited in their ability to fully characterize whether and how LLMs reason, despite being among the primary methods used in reasoning studies to date. For example, all the manually reviewed responses had relatively low inter-annotator agreement despite all reviewers being actively involved in the conceptualization of the error types. As a result, we only assessed the presence rather than the frequency of errors even though each model produced responses of varying length. Overall, our findings indicate that reasoning errors do occur in complex tasks; however, deeper analyses remain limited until advances in interpretability research provide greater insight into the underlying mechanics of LLM behavior.

**Study Limitations**

There are several limitations to the methods performed in this study. First, since the study results were limited to a single phenotype and dataset, specific insights related to the frequency and impact of reasoning errors may not necessarily generalize to other complex reasoning scenarios. Nevertheless, the results support a growing body of work indicating unfaithfulness in LLM responses. Second, the classification accuracies for all models were low. Since we had purposefully downgraded the quantization of all models to reduce confounding, we expected lower accuracies. However, as a result, we were unable to assess any relationship between low classification accuracy and presence of reasoning errors in this study. Third, due to model availability constraints, we were unable to use DeepSeek without Qwen distillation. While still considered a reasoning model, the Qwen-distilled version was not directly trained with reasoning-focused reinforcement learning and therefore, may exhibit different behaviors than the original DeepSeek model. Finally, there is no indication which, if any, of the experiments best correlate to the internal reasoning processes of each model.

**Future Directions**

There are several future directions to consider based on the results of this study. First, reasoning methods may be further improved by developing more robust interpretability techniques for the internal reasoning processes of LLMs. While prior work has explored various strategies to understand LLM reasoning,(32–34) important questions remain regarding the generalizability of these findings, the consistency of reasoning mechanisms across different LLMs, and the scalability and refinement of interpretability approaches. Another future direction is to explore optimization of CoT for computational phenotyping using the results of this study and additional reasoning analyses. While studies have already developed different methodologies for optimizing CoT prompts, optimal CoT, especially for computational phenotyping tasks, has not yet been achieved.(35–37) Finally, development of models specifically trained to avoid or correct reasoning errors may be necessary in cases where faulty reasoning either strongly impacts model performance or leads to otherwise unfavorable results. Recent studies have explored reinforcement,(22,38,39) self-supervised,(39,40) and contrastive learning techniques(41) for improving LLM reasoning capabilities with promising results.

## CONCLUSION

In this study, we outlined methods for exploring faulty reasoning errors within CoT responses for computational phenotyping tasks. In addition to enhancing PHEONA to incorporate reasoning experiments for evaluating reasoning errors in LLM-based applications to computational phenotyping, we provided empirical evidence of reasoning errors in a complex reasoning use case. These results underscore the need for further investigation into LLM interpretability for complex tasks and mitigation strategies for reasoning errors.

## FUNDING STATEMENT

This work was supported in part by the National Institute of General Medical Sciences of the National Institutes of Health under grant T32 GM132008.

## COMPETING INTERESTS

The authors declare no competing interests.

## ETHICS STATEMENT

Data were obtained through the eICU Collaborative Research Database, an open-source dataset with EHR data from ICUs across the United States (https://eicu-crd.mit.edu/). Credentialed access was obtained by completing necessary training steps and acknowledging the Data Use Agreement. No additional ethics committee or institutional review board approval was required for access and use of the data.

## CODE AVAILABILITY

All code from this study is available at the following repository:
https://github.com/spungit/SHRECandPHEONA

## FIGURES

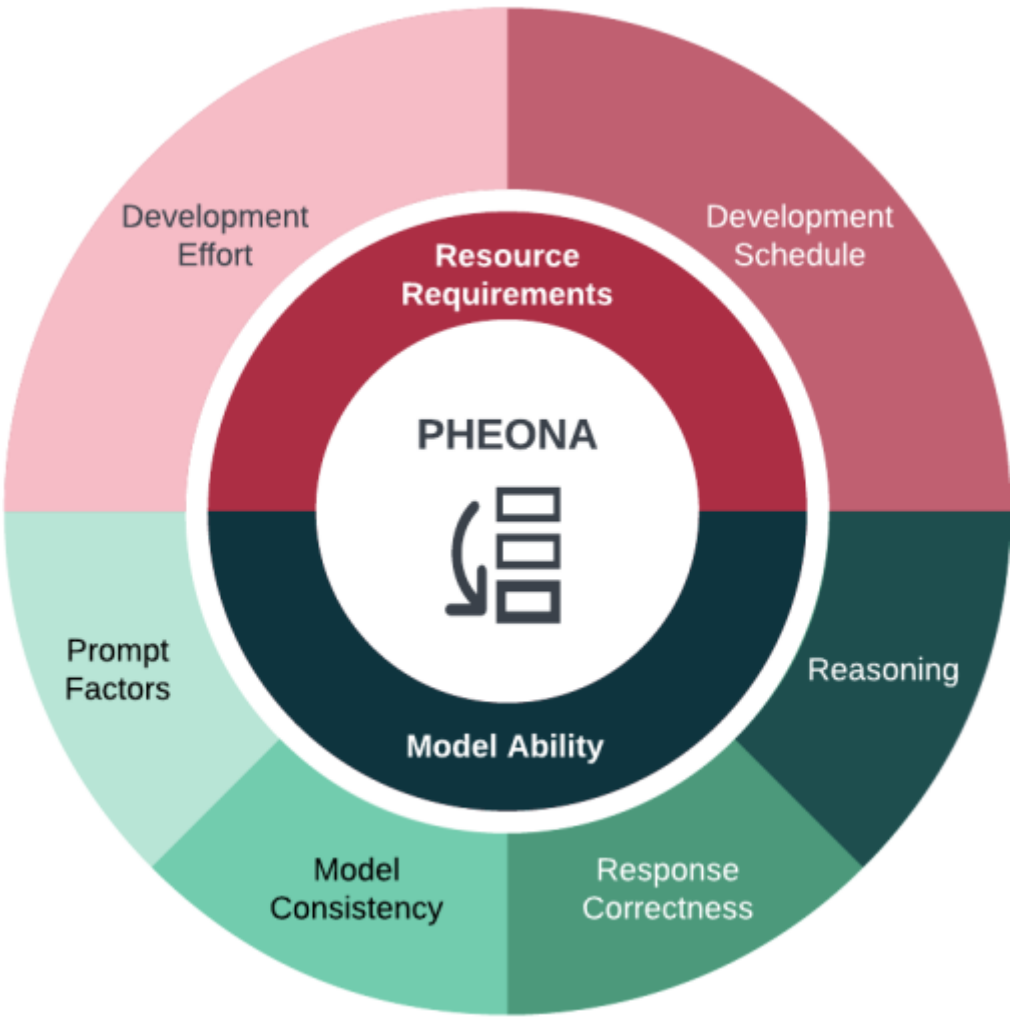

**Figure 1.** Overview of the components and related criteria of PHEONA (Evaluation of PHEnotyping for Observational Health Data), with an updated component, *Reasoning*, for evaluating faulty reasoning in model responses.

**Constructed Description:**
1: respFlowSettings: PEEP: 5
2: Airway: Intubated/oral ETT
3: Ventilation: Ventilated - with daily extubation evaluation
4: pulmonary|ventilation and oxygenation|mechanical ventilation
5: respFlowSettings: PEEP: 8
6: fentaNYL inj 50 mcg
7: respFlowSettings: PEEP: 10
8: midazolam (VERSED) 2 MG/2ML inj 1 mg
9: fentaNYL inj 25 mcg
10: respFlowSettings: PEEP: 12
11: propofol 1% inj
12: Propofol (ml/hr)
13: respFlowSettings: FiO2: 80
14: Airway: Intubated/trach-acute
15: pulmonary|surgery / incision and drainage of thorax|tracheostomy|percutaneous
16: pulmonary|surgery / incision and drainage of thorax|tracheostomy|performed during current admission for ventilatory support
17: respFlowSettings: PEEP: 14
18: respFlowSettings: Pressure Support: 5
19: respFlowSettings: Pressure Support: 7
20: respFlowSettings: TV/kg IBW: 6.8060

**Figure 2.** Example constructed description for a single encounter generated by ordering and concatenating previously classified concepts for acute respiratory failure respiratory support therapies and relevant medications.(5) In this example, we observe that the individual was intubated at Record 2 and remained on a ventilator for the remainder of the encounter.

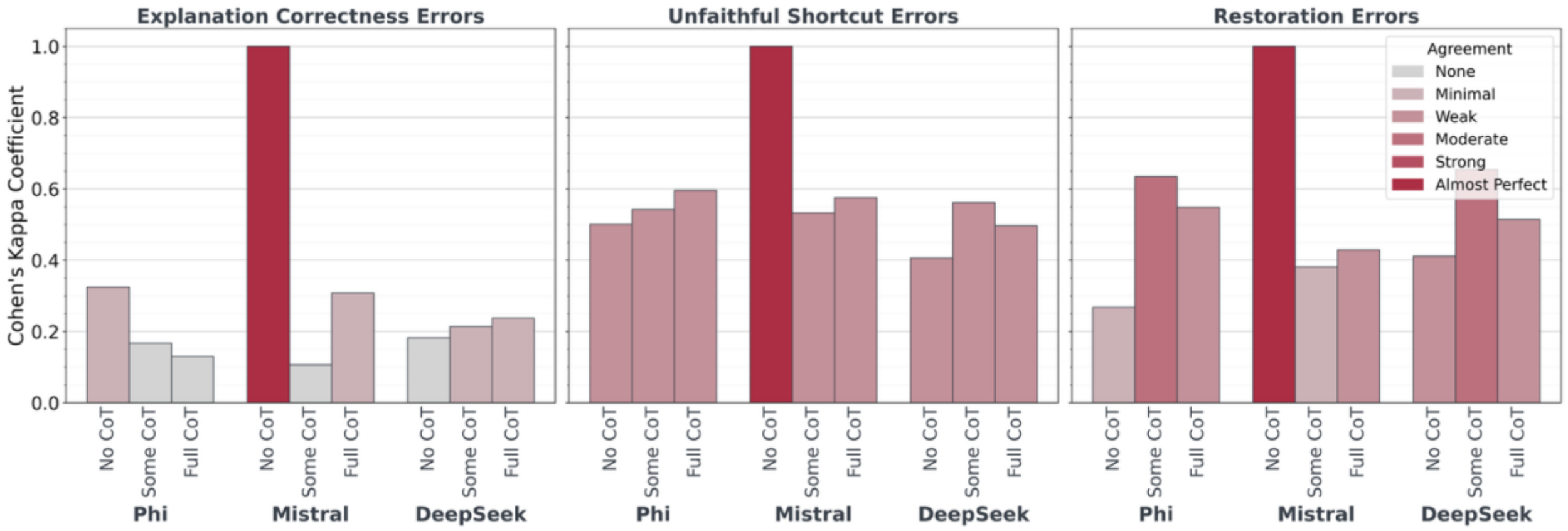

**Figure 3.** Inter-reviewer agreement calculated using Cohen's kappa coefficient for review of responses for explanation correctness, restoration, and unfaithful shortcut errors. Agreement levels were defined as follows: None (0 – 0.20), Minimal (0.21 – 0.39), Weak (0.40 – 0.59), Moderate (0.60 – 0.79), Strong (0.80 – 0.90), and Almost Perfect (Above 0.90).(23)

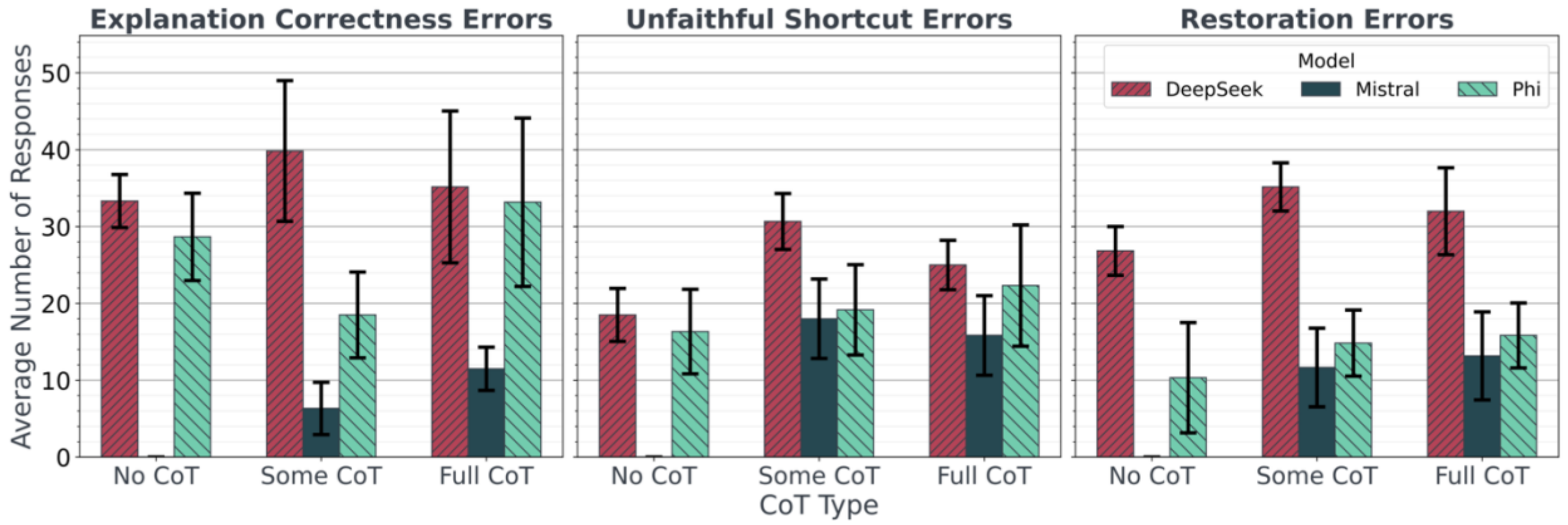

**Figure 4.** Average number of responses with explanation correctness, restoration, and unfaithful shortcut errors for each model and chain-of-thought (CoT) type. The number of responses with each error was determined for all 80 constructed descriptions across 6 total trials, 3 per reviewer. Error bars indicate the standard deviation of the number of responses with errors across all trials.

**Explanation Correctness Error Example: Phi**

**Relevant Description Records:**
*[Initial records]*
11: pulmonary|ventilation and oxygenation|mechanical ventilation
12: respFlowPtVentData: Peak Insp. Pressure: 11
13: Fentanyl (mcg/hr)
14: Propofol (Unknown)
15: respFlowPtVentData: Peak Insp. Pressure: 13
16: Propofol (mcg/kg/min)
17: respFlowSettings: Pressure Support: 5
18: respFlowPtVentData: Peak Insp. Pressure: 25
*[Remaining records]*

**Relevant Response:**
<OUTPUT>
SUMMARY:
Q1) SUMMARIZE RECORDS IN 3-5 SENTENCES.
A1) THE PATIENT'S ICU RECORDS SHOW INITIAL SETTINGS FOR POSITIVE END-EXPIRATORY PRESSURE (PEEP) AT 5 AND SUBSEQUENT ADMINISTRATION OF SEDATIVES, INCLUDING PROPOFOL AND MIDAZOLAM. **THE PATIENT WAS INTUBATED WITH AN ORAL ENDOTRACHEAL TUBE, AS INDICATED BY MULTIPLE RECORDS OF MECHANICAL VENTILATION AND PEAK INSPIRATORY PRESSURES OVER TIME.** SEDATION CONTINUED WITH FENTANYL ALSO BEING ADMINISTERED.

*[Remaining response.]*

</OUTPUT>

**Unfaithful Shortcut Error Example: Mistral**

**Relevant Description Records:**
1: respFlowSettings: PEEP/CPAP: 5
2: 100 ML VIAL : PROPOFOL 10 MG/ML IV EMUL
3: 1 ML CRTRDG-NDL : MORPHINE SULFATE 2 MG/ML IJ SOLN
4: respFlowSettings: PEEP/CPAP: 8
5: 5 ML VIAL : MIDAZOLAM HCL 1 MG/ML IJ SOLN
6: 2 ML VIAL : FENTANYL CITRATE 0.05 MG/ML IJ SOLN
7: respFlowCareData: FIO2 (%): 80
8: respFlowCareData: O2 Device: Ventilator
9: pulmonary|ventilation and oxygenation|mechanical ventilation
10: Airway: Intubated/oral ETT
11: Ventilation: Ventilated - with no daily extubation trial
*[Remaining records]*

**Relevant Response:**
<OUTPUT>
SUMMARY:

*[Description of records.]*

Q2) ARE REQUIRED MEDICATIONS PRESENT? ARE THERE >2 IMV/INTUBATION RECORDS? YES OR NO FOR IMV CRITERIA MET?
A2) YES, REQUIRED MEDICATIONS ARE PRESENT. MORPHINE SULFATE AND MIDAZOLAM HCL WERE ADMINISTERED (ORDER 3). **THERE ARE AT LEAST TWO IMV/INTUBATION RECORDS: "PULMONARY|VENTILATION AND OXYGENATION|MECHANICAL VENTILATION" (ORDER 9),** AND "AIRWAY: INTUBATED/ORAL ETT" (ORDER 10).

NIPPV CRITERIA:
Q3) ARE THERE >2 NIPPV RECORDS INDEPENDENT OF HFNI/NASAL CANNULA? YES OR NO FOR NIPPV CRITERIA MET?
**A3) NO. THERE ARE NO NIPPV RECORDS INDEPENDENT OF HFNI/NASAL CANNULA.**

*[Remaining response.]*

</OUTPUT>

**Restoration Error Example: DeepSeek**

**Relevant Description Records:**
*[Initial records]*
5: pulmonary|ventilation and oxygenation|oxygen therapy (< 40%)|nasal cannula
6: O2 Admin Device: hi flow
7: O2 Admin Device: hi-flow
8: respFlowSettings: FiO2: 80
9: DESCRIPTION: Source = Note; Concept = mask: mask
10: fentaNYL citrate (PF)
11: Oral ETT
*[Remaining records]*

**Relevant Response:**
<THINK>
*[Initial response setting up the problem.]*

**HFNI REQUIRES BOTH MEETING NIPPV CRITERIA AND AT LEAST ONE HFNI/NASAL CANNULA RECORD WITH HIGH FLOW. RECORDS 3,5,6,7 MENTION NASAL CANNULA OR HI-FLOW.** RECORD 8 SHOWS FIO2: 80, BUT WITHOUT THE FLOW RATE. IF THE FLOW IS ABOVE 15 L/MIN, IT WOULD BE HFNI. HOWEVER, SINCE WE DON'T HAVE EXPLICIT FLOW RATES, IT'S UNCLEAR.

*[Response related to other treatments.]*

- MULTIPLE MENTIONS OF NASAL CANNULA AND HI-FLOW DEVICES, POSSIBLY HFNI IF FLOW RATES ARE HIGH.
SINCE WE HAVE A SINGLE RECORD EACH FOR ETT AND MASK, **AND MULTIPLE FOR POSSIBLE HFNI WITHOUT CONFIRMED FLOW RATE, NONE OF THE TREATMENT CRITERIA ARE FULLY MET. THEREFORE, THE CLASSIFICATION IS NONE.**
</THINK>

**Figure 5.** Representative examples (each bolded and highlighted in orange) of explanation correctness, restoration, and unfaithful shortcut reasoning errors from phenotyping the constructed descriptions. For the explanation correctness error example, the model states that Invasive Mechanical Ventilation (IMV) was indicated by mechanical ventilation and peak inspiratory pressure records although these records are not exclusive to IMV and may be related to Noninvasive Positive Pressure Ventilation (NIPPV) or High-Flow Nasal Insufflation (HFNI). Similarly, for the unfaithful shortcut error example, the model classifies all ventilator records as IMV without acknowledging that several records (such as the ventilation settings and the mechanical ventilation record) could also indicate NIPPV or HFNI. For the restoration error

example, the model first states that there are multiple high-flow records and then later states that there are not enough high-flow records before classifying the record as None.

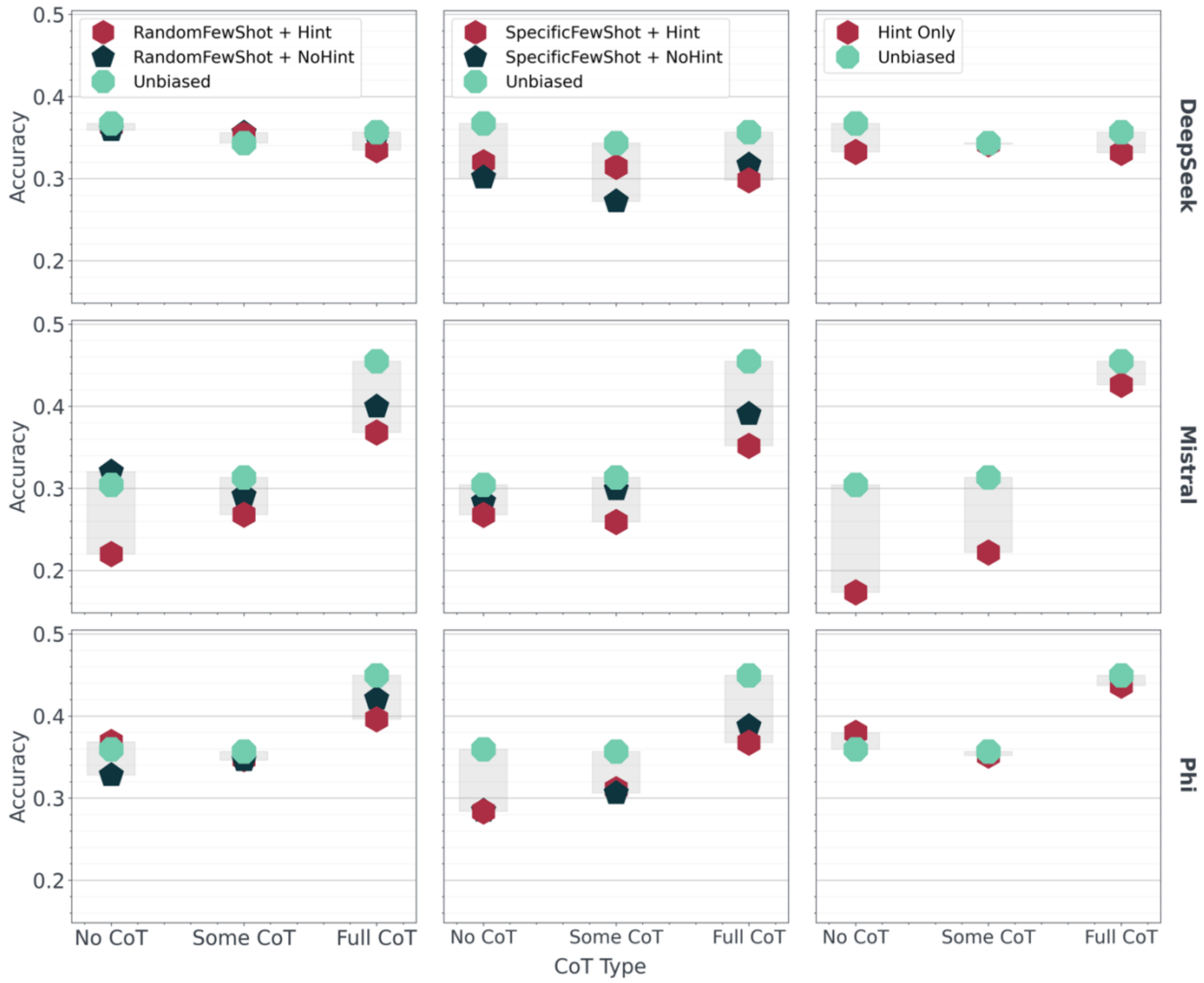

**Figure 6.** Average accuracy of responses for the *Few-Shot Biasing* and *Hint Biasing* experiments. Accuracies were calculated for 700 constructed descriptions (100 for each phenotype except IMV to HFNI) across 3 trials for each experiment, model, and chain-of-thought (CoT) type against phenotype ground truths.